\documentclass[twocolumn,prb,showpacs]{revtex4}
 \usepackage{graphicx}
\begin{document}

\title{Interaction of a Kink-Soliton with a Breather in a Fermi-Pasta-Ulam Chain}

\author{Ramaz Khomeriki \email{khomeriki@hotmail.com}
}

\affiliation{Department of Physics and Astronomy, The University of Oklahoma,
440 W. Brooks, Norman, OK 73019, USA;}

\affiliation{Department of Physics, Tbilisi State University, Chavchavadze ave.3, Tbilisi 380028, GEORGIA}

\date{\today}

\begin{abstract}

\noindent Collision process between breather and moving kink-soliton is investigated both
analytically and numerically in Fermi-Pasta-Ulam (FPU) chains. As it is shown by both
analytical and numerical consideration low amplitude breathers and soft kinks
retain their shapes after interaction. 
Low amplitude breather only changes the location after collision and remains static.
As the numerical simulations show, the shift of its position is proportional to the 
stiffness of the kink-soliton, what is in
accordance with the analytical predictions made in this paper. The numerical
experiments are also carried out for large amplitude breathers and some interesting effects
are observed: The odd parity large amplitude breather does not change position
when colliding with widely separated soft kink-antikink pair, whilst in the case of closely
placed kink-antikink pair the breather transforms into the moving one. 
Therefore it is suggested that the "harmless" objects similar to the kink-solitons 
in FPU chains could be used in order 
to displace or move the strongly localized structures in realistic
physical systems. In particular, the analogies with quasi-one dimensional 
easy-plane type spin structures are discussed.

\pacs{63.20.Ry; 05.45.Yv; 63.20.Pw}

\end{abstract}
\maketitle

\section{Introduction}

Chains of classical anharmonic oscillators can serve as models
for more complex physical systems which under definite conditions could be treated as one
dimensional objects, e.g., optical fibers, magnetic film waveguides, quasi-one dimensional
spin systems, DNA, ionic crystals, etc. Modeling various physical processes one can directly see
consequences using computer simulations and compare them with the established analytical schemes.
In the present paper it is proposed to model different nonlinear processes in chains of coupled
oscillators making simultaneous interpretations and predictions concerning real physical
systems.

Already the simplest model, as the one-dimensional chain of equal-mass oscillators
exhibits following nontrivial phenomena such as energy equipartition \cite{ruffo1},
\cite{lichtenberg1}, appearance of various
patterns \cite{ramaz} and localizations \cite{flach} (either moving \cite{flach2}
or static \cite{lepri}, \cite{lichtenberg2}),
different regimes of chaotic dynamics \cite{ruffo2}, \cite{politi}, etc. Therefore these
classical systems could serve as tools
for better understanding of nonlinear phenomena in completely different (on the first sight)
many-body systems. For instance,
invariance under the simple symmetry transformation $u_n\rightarrow u_n+ const$ ($u_n$
is a displacement of $n$-th oscillator) relates Fermi-Pasta-Ulam (FPU)
chain~\cite{fermi} (interparticle forces are functions of only relative displacements)
with a wide class of systems with continuous symmetries \cite{tribelsky},
e.g. quasi-one dimensional easy plane ferromagnets
and antiferromagnets \cite{lai}, ferrimagnetic spiral structures \cite{ramab}
and even quantum Hall double layer (pseudo) ferromagnets \cite{spielman}. Such
systems are characterized by the infinitely
degenerated energy ground state. Spontaneous breakdown of the symmetry (by choosing a
definite ground state) leads to the appearance of
gapless Goldstone mode forming the kink-solitons in low energy limit. These
localized solutions are well known for easy plane magnetic structures \cite{ramab}, \cite{nika}
and demonstrate similar properties as in the case of FPU chains
\cite{kosevich}, \cite{poggi}.

The main difference between Goldstone mode kink-solitons and ordinary kinks of the models
related to sine-Gordon equation (particularly, its discrete analogy -
Frenkel-Kantorova model \cite{braun}, \cite{savin}) is that the formers
do not carry topological charge. Besides that, the
kinks are believed to be the exact solutions \cite{duncan} in FPU chain.
Because of these circumstances it is expected that
they should not decay itself and do not destruct other localizations
during the scattering process as long as no energy
redistribution is required. In this connection it should
be mentioned that FPU chain, the linear spectrum of which is bounded from above,
exhibits another nontrivial solution in high energy limit. This solution represents
intrinsic localized mode (discrete breather) \cite{sievers}, \cite{page}, which in a low amplitude
limit could be considered as particular case of semi-discrete envelope soliton
\cite{hori}, \cite{konotop}. Let us note
a direct analogy of the above 
with quasi one-dimensional magnetic systems where similar localizations
have recently been discovered \cite{english1}, \cite{english2} or predicted \cite{salerno}.

As it follows from the analytical and numerical considerations made in the present paper
the kink-solitons are indeed "harmless": after interaction the shapes of both
kink and breather remain unchanged. The collision only causes the shift of 
the position of spatially localized breather or
its transformation into the slowly moving one.
In this connection let us make a comparison with strongly inelastic
scattering process between kinks and breathers of sine-Gordon equation 
\cite{malomed1}, \cite{malomed2}. Althogh, it should be mentioned that
in the latter case the nonlinear objects are solutions of continuous equation
unlike discrete FPU model considered in the present paper.

For analytical consideration in weakly nonlinear limit the multiple scale analysis 
will be used in order to present the quantitative picture for kink-breather collision. 
But firstly,
well known solutions for kink and breather will be briefly rederived in order to 
introduce the method of calculations \cite{oikawa}, \cite{ramazz}. 

\section{Analytical Solutions for Kink-Soliton and Breather in Weakly Nonlinear Limit}

The equations of motion of FPU oscillator chain are
\begin{equation}
\ddot u_n=(u_{n+1}-u_n)+(u_{n-1}-u_n)+(u_{n+1}-u_n)^3+(u_{n-1}-u_n)^3,
\label{fpu}
\end{equation}
where dots over $u_{n}$ express the time derivatives. Dimensionless units are
used so that the masses, the linear and nonlinear force constants and the
lattice spacing are taken equal to unity. The real displacements are expressed from
dimensionless ones ($u_n$) by dividing the latters
on the coefficient $\sqrt{K_4/m}$ where $m$ is a mass of particle and $K_4$ is a coefficient
before the anharmonic quartic term. Thus if the nonlinear interaction is strong enough it is
admissible to
have large values of $u_n$ (e.g. $u_n\gg 1$) and this does not cause the scattering of
neighboring particles.

Firstly let us derive the kink-soliton solution by assuming that $u_n$ smoothly varies
in space-time. Then it is appropriate to introduce slow variables
\begin{equation}
\xi_1=\varepsilon(n-v_1t); \qquad \tau_1=\varepsilon^3 t \label{1}
\end{equation}
and denote
\begin{equation}
u_n=\varphi_1(\xi_1,\tau_1), \label{2}
\end{equation}
where $\varepsilon$ is a formal small parameter indicating smallness or slowness of the
variables before which it appears. Substituting (\ref{2}) into the motion equation
(\ref{fpu}) and collecting the terms with the same order of $\varepsilon$ it becomes possible
to treat the problem perturbatively. In particular, the velocity $v_1$ is determined in
the second approximation over $\varepsilon$:
\begin{equation}
v_1=\pm1. \label{3}
\end{equation}
Without restricting of generality let us further consider the solution with negative velocity
$v_1=-1$. Other solution will be recovered simply changing the axis direction. Finally,
in the forth approximation over $\varepsilon$ the following nonlinear equation is obtained:
\begin{equation}
\frac{\partial^2\varphi_1}{\partial\xi_1\partial\tau_1}+\frac{1}{24}\frac{\partial^4\varphi_1}
{\partial\xi_1^4}+\frac{3}{2}\left(\frac{\partial\varphi_1}{\partial\xi_1}\right)^2
\frac{\partial^2\varphi_1}{\partial\xi_1^2}=0, \label{4}
\end{equation}
which is exactly integrable modified Korteweg-deVries equation \cite{ablowitz}
for the function $\partial\varphi_1/\partial\xi_1$. The equation (\ref{4})
was derived for FPU chain
in Refs. \cite{kosevich},\cite{poggi} and finally leads to the kink-like solution for
$u_n$:
\begin{equation}
u_n=\varphi_1=\sqrt{\frac{2}{3}}\left(arctan\left[e^{A\sqrt{6}\bigl(n+t+(A/2)^2t\bigr)}
\right]\right),
\label{5} \end{equation}
which has a similar form as the kinks for the sine-Gordon equation but note that although the tails
of the kink solution (\ref{5}) corresponds to the different ground states ($u_n=0$ and
$u_n=\pi/\sqrt{6}$ for $n\rightarrow-\infty$ and $n\rightarrow\infty$, respectively) these
ground states carry the same energy because of the mentioned symmetry $u_n\rightarrow
u_n+const$. These kinks do not carry topological charge and as far as they connect
degenerate ground states they could be called Goldstone mode kinks. It should be
also mentioned that in terms of relative displacements $v_n=u_{n+1}-u_n$ this object
is discretized version of Korteweg-deVries soliton and therefore the definition  of kink-soliton
is usually used in literature for its identification. The similar localized objects 
could be created in magnetic structures with
easy plane anisotropy where their appearance also is connected with the 
broken symmetry Goldstone mode. The
transverse component of such magnetic localization (in-plane component) has a kink like form, 
while out of easy plane component represents 
ordinary Korteweg-deVries soliton \cite{ramab}, \cite{nika}.

The solution (\ref{5})
is valid if one can neglect the higher derivatives. This could be achieved if the following
condition is satisfied for the kink stiffness:
\begin{equation}
6A^2\ll 1. \label{6}
\end{equation}

Afterwards let us rederive the breather solution using multiple scale analysis presenting
$u_n$ as multiplication of harmonic oscillation and smooth envelope function:
\begin{equation}
u_n=\frac{\varepsilon}{2}\varphi_2(\xi_2,\tau_2)e^{i(kn-\omega t)}+c.c. \label{7}
\end{equation}
where "c.c." denotes complex conjugation and new slow variables are defined as follows:
\begin{equation}
\xi_2=\varepsilon(n-v_2t); \qquad \tau_2=\varepsilon^2t. \label{8}
\end{equation}
As far as only small displacements are considered it is natural to neglect the higher harmonics
working in a rotating wave approximation.
Carrying out the procedure similar to the previous case (collecting terms with the same
harmonics and order of $\varepsilon$) in the first order over $\varepsilon$ well
known dispersion relation for linear excitations in FPU chain is obtained:
\begin{equation}
\omega=\omega_k\equiv\sqrt{2(1-cosk)}. \label{9}
\end{equation}
In the second approximation the expression for group velocity is derived:
\begin{equation}
v_2=\frac{sink}{\omega_k}\equiv\frac{d\omega_k}{dk}, \label{10}
\end{equation}
and finally we get the nonlinear Schr\"odinger equation for the envelope function $\varphi_2$
in the third approximation over $\varepsilon$:
\begin{equation}
i\frac{\partial\varphi_2}{\partial\tau_2}-\frac{\omega_k}{8}\frac{\partial^2\varphi_2}
{\partial\xi_2^2}-\frac{3}{8}\omega_k^3|\varphi_2|^2\varphi_2=0, \label{11}
\end{equation}
which permits bright soliton solution. Thus in terms of $u_n$ the envelope soliton solution
(moving with a group velocity $v_2$) is rederived (see e.g. Ref. \cite{hori}):
\begin{equation}
u_n=\frac{Bcos(nk-\tilde\omega_k t)}{ch\bigl[\sqrt{3/2}B\omega_k(n-v_2t)\bigr]}, \label{12} \end{equation}
$$ 
\tilde\omega_k=\omega_k\Bigl(1+(3/16)\omega_k^2B^2\Bigr), \qquad B\ll 1. 
$$
The breather solution is obtained by setting $v_2=0$, therefore
carrier wave number $k=\pi$ (thus $\omega=2$) should be considered
according to the relations (\ref{9}) and (\ref{10}).
Thus we get the expression for the low amplitude breather solution:
\begin{equation}
u_n=\frac{Bcos\bigl(\pi n-2t-(3/2)B^2t\bigr)}{ch\bigl(B\sqrt{6}n\bigr)}, \qquad B\ll 1 \label{13}
\end{equation}
which coincides with the corresponding breather solution obtained in Ref. \cite{lepri}.

\section{Interaction between Kink-Soliton and Breather}

\subsection{Analytical Results in Weakly Nonlinear Limit}

Now let us start the main task of the paper: analytical description of kink-breather
interaction. Keeping in mind that in absence of either kink or breather one should come
to the solutions (\ref{13}) or (\ref{5}), respectively, I am seeking for the solution in the 
following form (using again the rotating wave approximation):
\begin{equation}
u_n=\varphi_1(\xi_1,\tau_1)+\frac{\varepsilon}{2}\left[\varphi_2(\xi_2,\tau_2)
e^{i(\pi n-2t)+i\varepsilon \Omega(\xi_1,\tau_1)}+c.c.\right], 
\label{14}
\end{equation}
where the following choice for slow space-time variables is made:
$$
\xi_1=\varepsilon(n+t)-\varepsilon^2\Psi_1(\xi_2,\tau_2), \qquad \tau_1=\varepsilon^3 t;
$$
\begin{equation}
\xi_2=\varepsilon n-\varepsilon^2\Psi_2(\xi_1,\tau_1), \qquad \tau_2=\varepsilon^2 t.
\label{15} \end{equation}
Here the phase and argument shifts are introduced in order to decouple nonlinear equations.
Substituting (\ref{14}) into the initial equation of motion for FPU chain 
(\ref{fpu})
we get in the forth order over $\varepsilon$ for zero harmonic and in the third order
over $\varepsilon$ for the first harmonic the following two nonlinear equations: 
\begin{equation}
\frac{\partial^2\varphi_1}{\partial\xi_1\partial\tau_1}+\frac{1}{24}\frac{\partial^4\varphi_1}
{\partial\xi_1^4}+\frac{3}{2}\left(\frac{\partial\varphi_1}{\partial\xi_1}\right)^2
\frac{\partial^2\varphi_1}{\partial\xi_1^2}+
\label{16} \end{equation}
$$
+\frac{1}{2}\left[\frac{\partial^2\varphi_1}
{\partial\xi_1^2}+\frac{\partial\varphi_1}{\partial\xi_1}\frac{\partial}
{\partial\xi_2}\right]\left[\frac{\partial\Psi_1}{\partial\xi_2}+6|\varphi_2|^2\right]=0, 
$$
\begin{equation}
i\frac{\partial\varphi_2}{\partial\tau_2}-\frac{1}{4}\frac{\partial^2\varphi_2}
{\partial\xi_2^2}-3|\varphi_2|^2\varphi_2-\varphi_2\left[\frac{\partial\Omega}
{\partial\xi_1}+3\left(\frac{\partial\varphi_1}{\partial\xi_1}\right)^2\right]=0. \label{17}
\end{equation}
The variables $\xi_1$ and $\xi_2$ in Exp. (\ref{15}) are chosen such that
group velocities for noninteracting kink-soliton and breather (with carrier
wave number equal to $\pi$) are $v_1=-1$ and $v_2=0$, what itself guarantees the satisfaction
of the motion equation (\ref{fpu}) in the lower orders over
$\varepsilon$.

By letting
\begin{equation}
\frac{\partial\Psi_1}{\partial\xi_2}=-6|\varphi_2|^2, \qquad
\frac{\partial\Omega}
{\partial\xi_1}=-3\left(\frac{\partial\varphi_1}{\partial\xi_1}\right)^2 \label{18}
\end{equation}
we come again to the equations (\ref{4}) and (\ref{11}) for kink-soliton $\varphi_1$ and
breather $\varphi_2$ (with carrier wave number $k=\pi$).  The choice (\ref{18}) physically 
means that the interaction effects reduce only to the phase shifts of solitons
while the solitons' profiles remain unchanged in the leading approximation. 

Finally, in the fourth
approximation over $\varepsilon$ for the first harmonic the following equality is derived:
\begin{equation}
\frac{\partial\Psi_2}
{\partial\xi_1}=\frac{3}{2}\left(\frac{\partial\varphi_1}{\partial\xi_1}\right)^2. \label{19}
\end{equation}
According to the last relation breather acquires group velocity during the interaction
process, but as the kink passes it stops. The shift of breather's position could be calculated
from the following simple relations:
\begin{equation}
l_2=\int\limits_{-\infty}^\infty v_2dt=
\int\limits_{-\infty}^\infty \frac{\partial\Psi_2}
{\partial t}dt=\int\limits_{-\infty}^\infty \frac{\partial\Psi_2}
{\partial \xi_1}d\xi_1=\frac{\sqrt{6}}{2}|A|, \label{20}
\end{equation}
thus shift is always positive, i.e. breather will be shifted oppositely to the kink's
propagation direction (let us remind that group velocity of the kink has been chosen to be 
negative)
irrespective of the sign of $A$, therefore the shift is the same for both kink and antikink 
case. 

Denoting by $t_1$ and $t_2$ the times needed for kink to travel from the one side of the
chain to another in presence or absence of the breather on the way, one can calculate
the difference $\Delta t=t_2-t_1$ using similar to (\ref{20}) relations. Thus one gets:
\begin{equation}
\Delta t=2\sqrt{6}|B|. \label{21}
\end{equation}

The physical meaning of the expressions (\ref{20}) and (\ref{21}) could be simply understood
mentioning that in case of weakly nonlinear solitons' interaction the group velocities
of the solitons change only during the interaction process. Particularly, the breather
acquires the nonzero velocity while interacting with kink-soliton. Simultaneously, during
the same small time period the velocity of kink-soliton becomes larger than in case of its free
propagation. These circumstances cause the shift of low amplitude breather position and 
on the other hand the earlier arrival of the kink soliton at the left side of the chain
(see Fig. 1).

\begin{figure}[htp]
\begin{center}\leavevmode
\includegraphics[width=\linewidth]{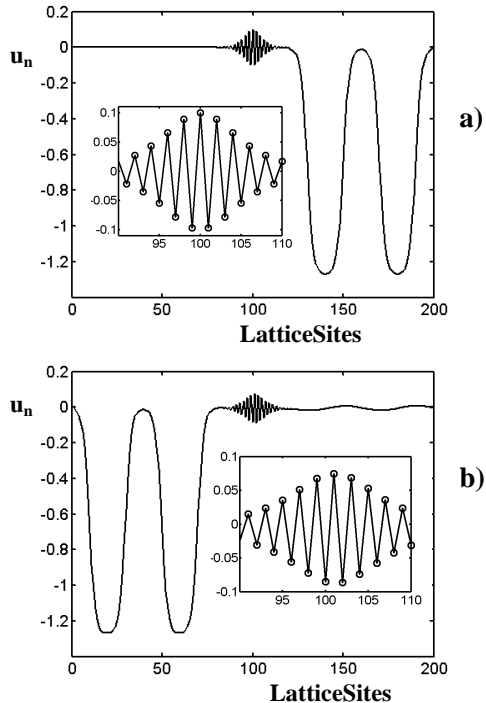}
\noindent
\vspace{-1cm}
\caption{Shapes and locations of kink-antikink pairs and low amplitude breather 
(a) before and (b) after collision. Kink-antikink pairs move from right to the 
left. Insets show the enlarged view of the breather. $u_n$ is the displacement of  $n$-th site.} \end{center}
\label{mamb.eps}\end{figure}

\subsection{Numerical Experiments}

As it was mentioned above, direction of the shift of the breather position 
does not depend on whether the kink or
antikink participate in the collision. Therefore in order to increase the interaction effect
two kink-antikink pairs are used for collision with breather. In the numerical experiment
the soft kink solution (\ref{5}) and low amplitude breather (\ref{13}) are put apart from 
each other in FPU chain with pinned boundary conditions (see Fig. 1). Numerical experiment
fully confirms analytical predictions: the nonlinear objects for which  
analitycal results (\ref{5}) and (\ref{13}) are valid [i.e. the conditions (\ref{6}) and 
(\ref{13}) are satisfied], behave in full accordance with formulas (\ref{20}) and (\ref{21}).
Particularly, as series of numerical experiments show, 
the shift of low amplitude breather position is proportional to the kink-soliton
stiffness and does not depend on its own amplitude. In Fig.1 the collision process between
two kink-antikink pairs with
the same stiffness $A=0.2$ and the breather with amplitude $B=0.1$ is expressed.
It is clear that the localized objects
behave as expected: they retain their shapes and low amplitude
breather changes its position and after interacton becomes again static. 
According to Fig. 1 the shift is equal to 
$l_2\approx 1$ as it expected from formula (\ref{20}), note only that as four localized
objects with the same stiffness are used (two kink-antikink pairs) the value obtained from 
the expression (\ref{20}) should be multiplied by the factor $4$. The interaction also causes
acceleration of the kink-antikink pair (propagation velocity of kink-antikink pair 
is larger during the interaction process in comparison with free propagation) as it follows 
from the formula (\ref{21}). 

\vspace{2cm}
\begin{figure}[ht]
\begin{center}\leavevmode

\includegraphics[height=6.5cm]{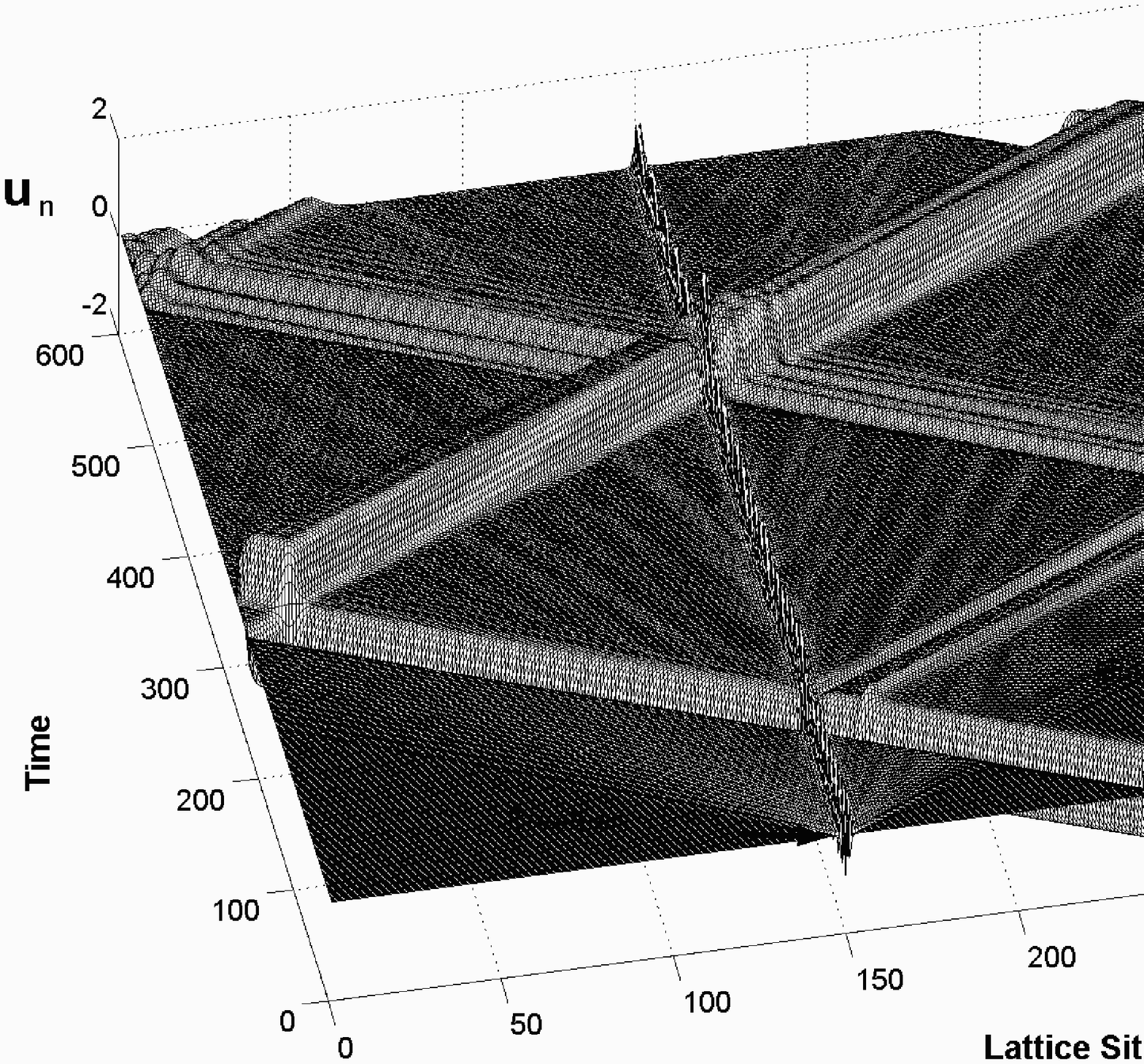}
\includegraphics[height=6.5cm]{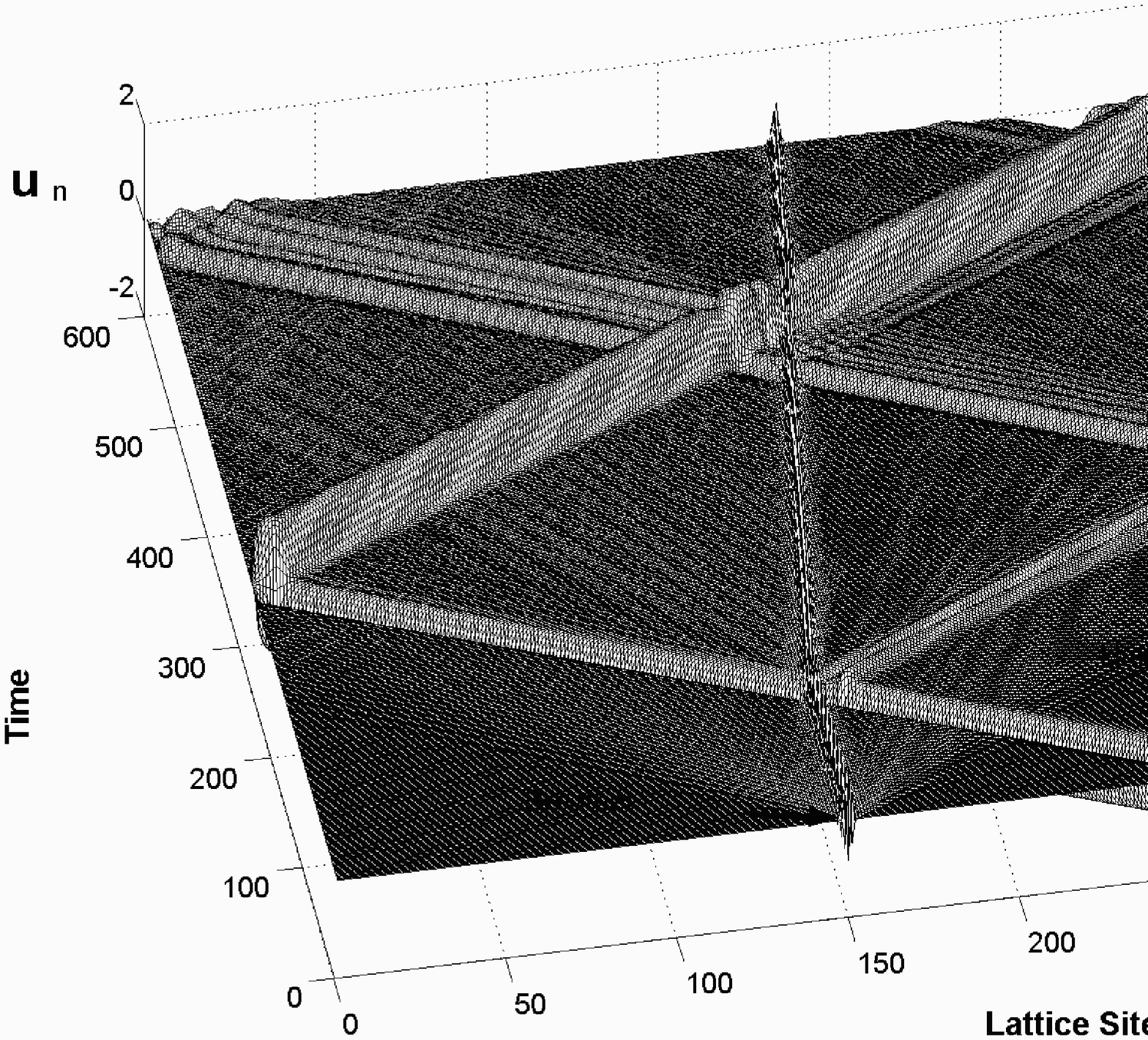}

\end{center}
\vspace{-2cm}
\caption{\noindent (a) collision of widely separated kink-antikink pair (stiffness $A=0.25$)
with strongly localized odd parity mode (amplitude $B=1$). 
(b) collision of the same breather with closely placed kink-antikink pair. Arrows 
show the initial position of the kink-antikink pair.}
\label{ppaml.eps}\end{figure}

Obviously, the weakly nonlinear approach fails considering large amplitude breather and/or
stiff kink-solitons. Therefore it is expected that the results could be different.
Indeed, the following phenomena are monitored when colliding soft kink-antikink pair 
with strongly localized odd parity mode. Breather with amplitude above
$B\approx 0.2$ does not change the position if soft
kink-solitons are used for collision. Moreover, breather does not change the position if widely
separated kink-solitons are used for collision. On the other hand, if there is a
relatively short distance 
(by order of inverse stiffness) between kink and antikink the breather starts moving after 
collision and does not stop. This behaviour is expressed in Fig. 2, where 
the strongly localazed odd parity mode (... 0, -1/2, 1, -1/2, 0 ...) is placed in the
center of FPU chain and different effects  are monitored for different separations 
between kink-antikink pairs (with stiffness $A=0.25$). Apparently this some kind of 
"quantum" phenomenon is caused by the fact 
that single soft kink (or antikink) does not carry enough impulse to transform the static
breather into the moving one, while the kink-antikink pair is able to do so. This statement 
is consistent with the observation of chaotic breathers (see e.g. Ref. \cite{ruffo1}) and 
with the lattice 
quantization procedure recently proposed in Ref. \cite{konotop}. It should be 
mentioned that all localized objects retain their shapes after interaction and the acceleration
of kink-antikink pair is still observed during the interaction process.

\vspace{-3cm}
\begin{figure}[htp]
\begin{center}\leavevmode
\includegraphics[width=\linewidth]{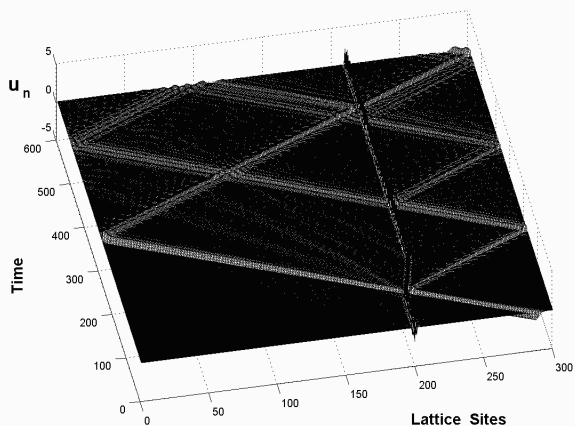}
\noindent
\vspace{-3cm}
\caption{Collision of closely placed kink-antikink pair (stiffness $A=0.35$)
with larger amplitude (B=2) odd parity mode. Note that although the breather starts
to move after collision (as in the previous figure), further it is trapped by
the lattice.}
\end{center}
\label{paml.eps}\end{figure}

Considering the breathers with larger amplitudes, naturally, the stiffer kink-antikink pairs
will be required to displace the breather. A number of numerical experiments have been made
for breathers with larger amplitudes. The general feature is represented in Fig. 3, 
where the collision of
odd parity mode (..., 0, -1, 2, -1, 0, ...) and kink-antikink pair with stiffness $A=0.35$
is demonstrated.
Like the previous case the breather does not change the position colliding with widely
separated kink-antikink pair. However, in contrast to the previous case, although the breather
starts to move because of the collision with closely placed kink-antikink pair, after some time
it is trapped by the lattice sites and stops. 

It should be mentioned that according to the analytical results and numerical experiments
the low amplitude breather acquires the group velocity only during the interaction process,
while the large amplitude breather once starting to move does not stop (for intermediate
amplitudes) or it will further trapped by the lattice sites (for larger amplitudes,
approximately $B>1.4$).

In the present paper only the results concerning collision process with odd parity mode 
are presented because, as numerical simulations show \cite{sievers}, the even parity mode
spontaneously starts to move and decays after finite time period, while odd parity
mode remains well localized in the absence of collisions with other nonlinear objects. 
At the same time only soft kink-solitons are the subject of the study because stiff kinks
sufficiently perturb background and it is hard to see what causes displacement or 
pinning of the breather.

Let us mention also that reflected kink-antikink pairs almost return the low amplitude breather 
to its initial position, while they are unable to change the picture in case of large
amplitude odd parity mode. As it is seen form Fig. 2 moving odd parity mode with amplitude
B=1 does not react
on the collision with reflected kink-antikink pair and remains moving with the same velocity.
On the other hand the reflected kink-antikink pair can not displace (see Fig. 3)
larger amplitude breather (B=2). 

\section{Conclusions}

Summarizing, as it is shown above, kink-solitons could be used to
displace or move the static localization without the destruction of the latter. 
The direct analogy
with quasi-one dimensional easy plane magnetic structures should be quoted again. 
The existence of
continuous U(1) symmetry in magnetization vector space for 
this systems allows the presence of the broken symmetry
Goldstone mode which because of the existing nonlinearity forms
kink-soliton. These large wavelength nonlinear excitations have been studied for easy plane
antiferromagnets in presence of applied magnetic field along anisotropic axis \cite{nika}
and in spiral structures \cite{ramab}. On the other hand it is known \cite{lai} that
the band edge excitations in easy plane type antiferromagnets form stable intrinsic localized
spin wave modes (ILSM) having odd parity structure in large amplitude limit.
Thus, a new theoretical study and corresponding realistic experiments could be planned 
in order to investigate and observe the effects 
caused by the interaction between magnetization kink-solitons and ILSM in quasi-one 
dimensional easy plane structures. As the analogy is almost
straightforward it could be predicted that kink-solitons in the easy-plane magnetic
structures should cause the same effect as in FPU chains. Particularly, they can displace
or move the strongly localized objects without their destruction.

The author is very thankful to Lasha Tkeshelashvili for the suggestions. The discussions with 
Stefano Lepri and Stefano Ruffo are also greatly acknowledged. The work was made possible in part 
by the NSF-NATO award No DGE-0075191.


\begin{thebibliography}{99}

\bibitem{ruffo1} T. Cretegny, T. Dauxois, S. Ruffo, A. Torcini, Physica D,
{\bf 121}, 109, (1998).
\bibitem{lichtenberg1} K. Ullmann, A.J. Lichtenberg, G. Corso, Phys. Rev. E, 
{\bf 61}, 2471, (2000).
\bibitem{ramaz} R. Khomeriki, S. Lepri, S. Ruffo, Phys. Rev. E, {\bf 64}, 056606, (2001).
\bibitem{flach} S. Flach, C.R. Willis, Phys. Rep., {\bf 295}, 181, (1998).
\bibitem{lepri} Yu. A. Kosevich, S. Lepri, Phys. Rev. B, {\bf 61 }, 299
(2000).
\bibitem{lichtenberg2} J. De Luca, A. J. Lichtenberg, S. Ruffo, Phys. Rev. E, {\bf 60}, 3781,
(1999).
\bibitem{flach2} S. Flach, Y. Zolotaryuk, K. Kladko, Phys. Rev. E., {\bf 59}, 6105, (1999).
\bibitem{ruffo2} V. Latora, A. Rapisarda, S. Ruffo, Phys. Rev. Lett., {\bf 80}, 692, (1998).
\bibitem{politi} A. Pikovsky, A. Politi, Phys. Rev. E, {\bf 63}, 036207, (2001).
\bibitem{fermi} E. Fermi, J. Pasta, S. Ulam, M. Tsingou, in {\it The
Many-Body Problems}, edited by D.C. Mattis (World Scientific,
Singapore, 1993 reprinted).
\bibitem{tribelsky} M.I. Tribelsky, Uspekhi Fizicheskikh Nauk, {\bf 167} (2), 167, (1997);
Physics-Uspekhi, {\bf 40} (2), 159, (1997).
\bibitem{lai} R. Lai, A.J. Sievers, Phys. Reports, {\bf 314}, 147, (1999).
\bibitem{ramab} N. Giorgadze, R. Khomeriki, Physica B, {\bf 252}, 274, (1998).
\bibitem{spielman} I. B. Spielman, J. P. Eisenstein, L. N. Pfeiffer, K. W. West, Phys. Rev. Lett.,
{\bf 87}, 036803 (2001).
\bibitem{nika} E.B. Volzhan, N.P. Giorgadze, A.D. Pataraya, Sov. Phys. Solid State, 
{\bf 18}, 1487, (1976).
\bibitem{kosevich} Yu. A. Kosevich, Phys. Rev. B, {\bf 47}, 3138, (1993).
\bibitem{poggi} P. Poggi, S. Ruffo, H. Kantz, Phys. Rev. E, {\bf 52}, 307, (1995).
\bibitem{braun} O. M. Braun, Yu.A. Kivshar, Phys. Rep. {\bf 306}, 1, (1998).
\bibitem{savin} A.V. Savin, Zh. Eksp. Teor. Fiz., {\bf 108}, 1105, (1995).
\bibitem{duncan} D.B. Duncan, J.C. Eilbeck, H. Feddersen, J.A.D. Wattis, Physica D, {\bf 68},
1, (1998).
\bibitem{sievers} A.J. Sievers, S. Takeno, Phys. Rev. Lett., {\bf 61}, 970, (1988).
\bibitem{page} J.B. Page, Phys. Rev. B, {\bf 41}, 7835, (1990).
\bibitem{hori} K. Hori, S. Takeno, J. Phys. Soc. Japan, {\bf 61}, 4263, (1992).
\bibitem{konotop} V.V. Konotop, S. Takeno, Phys. Rev. E, {\bf 63}, 066606, (2001).
\bibitem{english1} U.T. Schwarz, L.Q. English, A.J. Sievers, Phys. Rev. Lett., {\bf 83},
223, (1999). 
\bibitem{english2} L.Q. English, M. Sato, A.J. Sievers, J. Appl. Phys. {\bf 89}, 6707, (2001). 
\bibitem{salerno} V.V. Konotop, M. Salerno, S. Takeno, Phys. Rev. B, {\bf 58}, 14892, (1998).
\bibitem{malomed1} B.A. Malomed, Physica D, {\bf 15}, 374, (1985) ; {\bf 15} 385 (1985).
\bibitem{malomed2} Yu.S. Kivshar, B.A. Malomed, Physica D, {\bf 24}, 125, (1987).
\bibitem{oikawa} M. Oikawa, N. Yajima, J. Phys. Soc. Japan, {\bf 37}, 486, (1974).
\bibitem{ramazz} N. Giorgadze, R. Khomeriki, Phys. Rev. B, {\bf 60}, 1247, (1999). 
\bibitem{ablowitz} M.J. Ablowitz, P.A. Clarkson, {\it Solitons, Nonlinear Evolution Equations
and Inverse Scattering}, Cambridge University Press, Cambridge, (1991).

\end{thebibliography}
\end{document}